\documentclass[preprint,jcp,floats,amsmath,amssymb]{revtex4}

\usepackage{amsmath, amsthm, amssymb,graphicx,mathtools,algpseudocode,algorithmicx,bbm,color}
\newcommand{\x}{{\bf x}}

\begin{document}
\title{The Linear-Noise Approximation and moment-closure approximations for stochastic chemical kinetics}
\author{A Singh}
\affiliation{Department of Electrical and Computer Engineering, and Department of Biomedical Engineering,  University of Delaware, Newark, DE, USA}

\author{R Grima}
\affiliation{School of Biological Sciences, University of Edinburgh, Mayfield Road, Edinburgh EH9 3JR, Scotland, UK}

\begin{abstract}
This is a short review of two common approximations in stochastic chemical and biochemical kinetics. It will appear as Chapter 6 in the book ``Quantitative Biology: Theory, Computational Methods and Examples of Models" edited by Brian Munsky, Lev Tsimring and Bill Hlavacek (to be published in late 2017/2018 by MIT Press). All chapter references in this article refer to chapters in the aforementioned book. 
\end{abstract}
\maketitle

\vspace{0.8cm}

As discussed in the previous chapter, the time between successive chemical reaction events is random leading to fluctuations (noise) in the number of molecules \cite{Raj:2008Cell,eldar2010,Raser:2005Science,Singh:2012MSB,Dar:2012PNAS}. This noise is particularly evident when the average number of molecules is small as is often the case within individual cells.  Increasing evidence suggests that random fluctuations (noise)\index{noise} in protein copy numbers play important functional roles, such as driving genetically identical cells to different cell fates \cite{Balazsi:2011,Norman:2013Nature,Losick:2008Science,Singh:2009COM}. Moreover, many diseased states have been attributed to elevated noise levels in specific proteins \cite{Bahar:2006Nature,Brock:2009NatureReviewsGenetics}. 
The dynamics of biomolecular circuits, assuming well-mixed conditions, is mathematically described using the Chemical Master Equation (CME) framework (see previous chapter). Unfortunately, because the CME\index{chemical master equation} is not solvable for most chemical processes, stochastic analysis of biochemical systems relies heavily on Monte Carlo simulation techniques (\cite{gillespie2003,Rathinam:2003JCHP,Anderson:2007} (see Chapter 7).  These come at a significant computational cost due to the extensive ensemble averaging needed to obtain statistically significant results. In addition, these techniques do not provide any closed-form solutions that enable a systematic understanding of how noise is regulated in biochemical systems. 

\section{Introduction}

 In this chapter, the stochastic dynamics of these models is investigated through computation of lower-order statistical moments\index{moment} (such as the mean and the variance) for the copy numbers. We illustrate how differential equations describing the time evolution of moments can be derived from the CME.  In most cases, these equations cannot be solved numerically as they are not closed, in the sense that the time derivative of the lower-order moments depends on higher-order moments. State-of-the-art moment closure\index{moment! closure} schemes, that close these equations by expressing higher-order moments as approximate nonlinear functions of lower-order moments, are reviewed.  Finally, we discuss the Linear Noise Approximation (LNA)\index{linear! noise approximation}, an alternative powerful method for obtaining moments which is based on a small noise approximation of the CME.  These methods, which lead to ordinary differential equations for the approximate moments, offer a computationally tractable alternative to Kinetic Monte Carlo methods\index{Kinetic Monte Carlo}. Both closure techniques and LNA are illustrated on examples drawn from gene expression and enzyme kinetics. Towards the end, we briefly discuss software packages implementing the above methods for computing moment dynamics starting from a given description of biochemical reactions. 

\section{Stochastic Models of Biochemical Systems}
In this section, we review some background on stochastic modeling of biochemical reaction networks. Consider a well-mixed system of $N$ species ${S}_i$, $i \in \{1,\ldots, N\}$ that interact through $M$ reactions 
${R}_j$, $j \in \{1,\ldots, M\}$. In the context of cellular processes, species denote biological entities such as genes, RNAs, proteins, etc. Let $x_i (t)$ be the number of molecules of species $S_i$ at time $t$. At high population counts, the time evolution of the state vector $\x(t)=(x_1(t),\ldots,x_N(t))^T$ can be treated as a continuous and deterministic process governed by ordinary differential equations (see Chapter 3), often referred to as \emph{chemical rate equations} \cite{Wilkinson:1980VNRC}. However, this deterministic framework is not a very useful description within single cells where many species occur at very low molecular counts. The time evolution of such low-copy biochemical species is more accurately represented by a stochastic formulation of chemical kinetics which treats $\x(t)$ as a continuous-time, discrete-state Markov process \cite{Gillespie:1976,McQuarrie}.

In the stochastic formulation, molecular counts are positive integers that change by discrete jumps whenever a reaction occurs. The goal of this approach is to determine the joint probability density function (pdf) of the population count $\x(t)$, which evolves according to a Forward Kolmogorov Equation, often referred to as the\index{chemical master equation} \emph{Chemical Master Equation} (CME) \cite{Gillespie:1976,McQuarrie,Wilkinson:2011Chapman,Wilkinson:2011Chapman}. This equation is defined as follows. Let each reaction $R_{j}$ be assigned a probability $w_{j}(\x)dt$ that it will occur somewhere in the compartment volume in the next  ``infinitesimal'' time interval $(t,t+dt)$, where the \emph {propensity function}  $w_{j}(\x)$ is a polynomial in the population count (assuming mass-action kinetics) with units of inverse time \cite{Gillespie:1976}. By the laws of probability, it is easy to deduce the CME for the time-evolution equation for the probability $P(\x,t)$ that the system is in state $\x$ at time $t$: 
\begin{equation} 
\label{CMEgen}
\frac{dP(\x,t)}{dt} = \sum_{j=1}^M (w_{j}(\x - {\boldmath{v}}_j) P(\x - {\boldmath{v}}_j,t) - w_j(\x) P(\x,t)),
\end{equation}
where $\boldmath{v}_j$ is the $j^{th}$ column vector of the stoichiometric matrix $\boldmath{S}$ whose $i-j^{th}$ element $S_{ij}$ is the change in the number of molecules of species ${S}_i$ when reaction $R_j$ occurs. 

\begin{figure}[!t]
  \label{11}
  \centering
\includegraphics[width=7in]{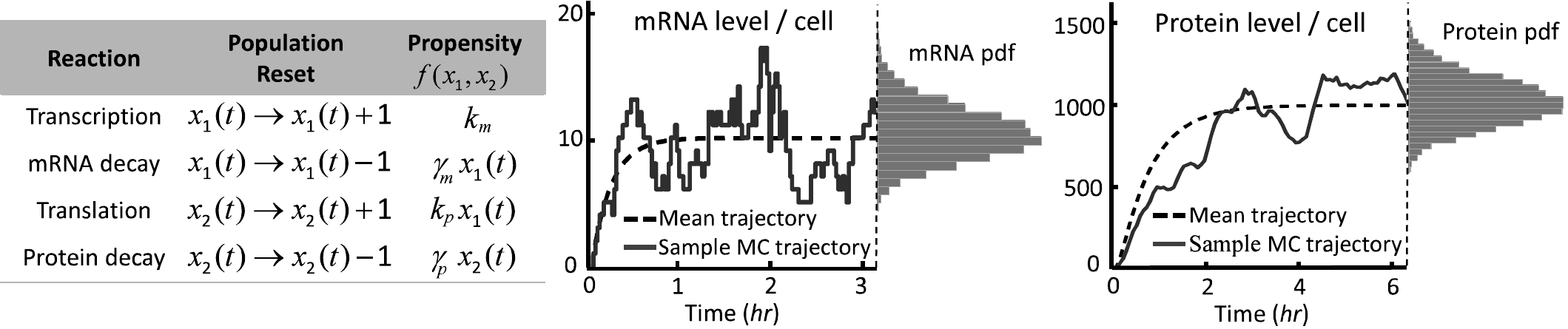}
   \caption{\small {\bf Illustration of a two-stage model of stochastic gene expression}. Table illustrating the model where stochastic occurrences of transcription, translation and degradation reactions increase/decrease protein ($x_2$) and mRNA ($x_1)$ counts by one (second column in the table). The third column lists the propensity function $w(x_1,x_2)$, which determines how often the reactions occur. Protein/mRNA trajectories corresponding to a Monte Carlo (MC) simulation run are shown for a certain parameter set together with the mean trends and the steady-state histograms obtained from $10,000$ Monte Carlo simulations performed using the Stochastic Simulation Algorithm (SSA) \cite{Gillespie:1976}.}\label{SinghFig1}
 \end{figure}

As an example, consider a highly simplified two-stage model of gene expression that can be represented by the following reactions:
\begin{equation} \label{re0}
\begin{aligned}
& Gene \xrightarrow{~k_m~} Gene + mRNA, \quad mRNA \xrightarrow{~\gamma_m~} \emptyset, \\
& mRNA \xrightarrow{~k_p~} mRNA + Protein,\quad Protein \xrightarrow{~\gamma_p~} \emptyset,
\end{aligned}
\end{equation}
where $k_m$ is the rate at which mRNAs are transcribed from an active gene. Proteins are produced from each mRNA at a translation rate $k_p$. Further, mRNAs and proteins degrade at constant rates $\gamma_m$ and $\gamma_p$, respectively. Based on the above formulation, the stochastic model for gene expression is illustrated in Fig. \ref{SinghFig1} together with sample realizations of mRNA/proteins levels obtained using Kinetic Monte Carlo simulations\index{Kinetic Monte Carlo}.
 
For the stochastic gene expression model, the CME is given by:
\begin{equation*}\label{CME}
\begin{aligned}
\frac{dP (x_1,x_2,t)}{dt}= &
 {k}_{m}P(x_1-1,x_2,t)+ {\gamma}_{m}(x_1+1)P(x_1+1,x_2,t)+ k_p x_1 P(x_1,x_2-1,t)+ \\ &{\gamma}_{p}(x_2+1)P(x_1,x_2+1,t) -P(x_1,x_2,t)( {k}_{m}+{\gamma}_{m} x_1+k_p x_1+{\gamma}_{p} x_2),
 \end{aligned}
\end{equation*}
where $P(x_1,x_2,t)$ denotes the probability of observing $x_1$ molecules of mRNA and $x_2$ molecules of protein at time $t$. It turns out that even for such simple stochastic models, the CME is difficult to solve exactly \cite{Shahrezaei:2008PNAS}. Typically, the pdf is computed numerically through the Finite State Projection Algorithm \cite{Munsky:2006JCP} or various Monte Carlo techniques \cite{gillespie2003,Rathinam:2003JCHP,gillespie2001,gibson-bruck,Cao:2004,Anderson:2007} at a significant computational cost. {\em Since one is often interested in computing only the lower-order moments for the number of molecules of the different species involved (for example, means, variances, correlations, skewnesses, etc.), much time and effort can be saved by directly computing these moments without actually having to solve for the pdf or by tedious ensemble averaging of Kinetic Monte Carlo simulations}. Next, we describe how differential equations for the time evolution of statistical moments can be derived from the CME and highlight challenges in solving these equations for nonlinear systems.

\section{Time Evolution of Statistical Moments}
\index{moment}

Given a vector $\{l_1,l_2,\ldots,l_N\}$ of
$N$ non-negative integers, the
\emph{uncentered} statistical moment of the population counts $\x=(x_1,\ldots,x_N)^T$  is defined as $\left \langle{x}_1^{l_1}{x}_2^{l_2} \cdots {x}_N^{l_N}\right \rangle$, where $\langle . \rangle$ stands for the expected value. We refer to the sum $\sum_{i=1}^N{l_{i}}$  as the  \emph{order of the
moment}. It is straightforward to show starting from the CME \eqref{CMEgen} that the time derivative of an uncentered moment of the population count is given by:
\begin{equation}\label{dyn}
\small
\frac{d \left \langle{x}_1^{l_1}{x}_2^{l_2} \cdots {x}_N^{l_N}\right \rangle}{dt}=\left \langle \sum_{j=1}^M w_j(\x)\left\{\left[\prod_{i=1}^N({x_i+S_{ij}})^{l_{i}}\right]-{x}_1^{l_1}{x}_2^{l_2} \cdots {x}_N^{l_N}\right\}\right \rangle,
\end{equation}
where $w_j(\x)$ is the propensity function of reaction ${R}_j$, and $x_i+S_{ij}$ is the population count of species ${S_i}$ when reaction ${R}_j$ occurs \cite{Hespanha:2005Control,Singh:2011IEEETAC,Singh:2010PTRSA,Grima:2012JCHP,Singh:2013PLOSONE}. Since the propensity functions are polynomials in the population count (according to the law of mass action), it follows from \eqref{dyn} that the time derivative of an uncentered moment $\left \langle{x}_1^{l_1}{x}_2^{l_2} \cdots {x}_N^{l_N}\right \rangle$ is a linear combination of uncentered moments of $\x=(x_1,\ldots,x_N)^T$. If all reactions have \emph{linear propensity functions} $w_j(\x)$, then the time derivative of a $l^{th}$ order moment is a linear combination of moments of order up to $l$  \cite{Singh:2011IEEETAC}.
We illustrate this point using the two-stage gene expression model in Fig.\ \ref{SinghFig1}. Using the propensity functions in the table in Fig.\ \ref{SinghFig1}, \eqref{dyn} reduces to:
\begin{equation}
        \begin{aligned}
 \frac{d\langle x_1^{l_1}x_2^{l_2} \rangle}{dt}=&\left \langle k_m \left( (x_1+1)^{l_1}x_2^{l_2}-x_1^{l_1}x_2^{l_2}\right)\right \rangle + \left \langle \gamma_m x_1\left( (x_1-1)^{l_1}x_2^{l_2}-x_1^{l_1}x_2^{l_2}\right)\right \rangle\\
&\left \langle k_p x_1 \left( x_1^{l_1}(x_2+1)^{l_2}-x_1^{l_1}x_2^{l_2}\right)\right \rangle + \left \langle \gamma_p x_2\left( x_1^{l_1}(x_2-1)^{l_2}-x_1^{l_1}x_2^{l_2}\right)\right \rangle.
  \label{dynnf}
        \end{aligned}
\end{equation}
Recall that $x_1(t)$ and $x_2(t)$ denote the single-cell level of the mRNA and protein, respectively. Time evolution of the mean protein level is simply obtained by setting $l_1=0$ and $l_2=1$ in \eqref{dynnf}. By appropriately choosing $l_1$ and $l_2$,  the first- and second-order moment dynamics can be derived as: 
\begin{subequations}\label{2d}
\begin{align}
&\frac{d\langle x_1 \rangle}{dt}= k_{m}- \gamma_{m}\langle x_1  \rangle, \\
&\frac{d\langle x_2 \rangle}{dt}= k_{p}\langle x_1 \rangle  -\gamma_p \langle x_2 \rangle,\\
&\frac{d\langle x_1 x_2 \rangle}{dt}=
k_{m} \langle x_2 \rangle+ k_{p} \langle x_1^2 \rangle - \gamma_{m}\langle x_1 x_2 \rangle- \gamma_{p}\langle x_1 x_2 \rangle ,\\
&\frac{d\langle x_1^2 \rangle}{dt}=k_m + 2  k_m \langle x_1 \rangle + \gamma_m\langle x_1 \rangle -  2 \gamma_m\langle x_1^2 \rangle, \\
&\frac{d\langle x_2^2 \rangle}{dt}=k_p\langle x_1 \rangle+ 2 k_p\langle x_1x_2 \rangle  + \gamma_p\langle x_2 \rangle- 2 \gamma_p\langle x_2^2 \rangle.
\end{align}
\label{moment equations}
\end{subequations}
Note that these equations depend on moments up to order two. The steady-state moments can be obtained by setting the time derivatives in \eqref{2d} to zero which leads to:
\begin{equation}\label{ssp}
\langle x_1 \rangle=   \frac{k_m }{\gamma_m}, \ \ \langle x_2 \rangle=   \frac{k_p \langle x_1 \rangle}{\gamma_p}, \ \ \langle x_1^2 \rangle-\langle x_1 \rangle^2=   \frac{k_m }{\gamma_m}, \ \ \langle x_2^2 \rangle-\langle x_2 \rangle^2=\frac{k_p}{\gamma_p}\frac{k_m }{\gamma_m}\left(1+\frac{k_p}{\gamma_m+\gamma_p} \right).
\end{equation}
The last equation in \eqref{ssp} provides important insights into how variance in protein levels scales with the transcription ($k_m$) and translation rates ($k_p$), consistent with experimental observations \cite{Newman:2006NatureGenetics,Bar-Even:2006NatureGenetics,Ozbudak:2002}. In general, for linear propensity functions, the time evolution of a vector $\boldmath{\mu}$ consisting of all the moments up to order ${{L}}$ of $\x$ is given by a linear dynamical system compactly written as:
\begin{equation}\label{eq:linearsys}
\small
\frac{d{\boldmath{\mu}}}{dt}={ a}+A {{{ \boldmath{\mu}}}},
\end{equation}
where vector ${a}$ and matrix $A$ depend on the model parameters. {\em Thus, for reactions with linear propensity functions (i.e., zero and first-order reactions), moments can be obtained exactly by solving \eqref{eq:linearsys}}.

Next, consider the scenario where reactions have \emph{nonlinear propensity functions},\index{nonlinear! reactions} in particular, polynomial functions of order 2 and higher (which often represent either bimolecular reactions or effective reactions which lump together a number of simpler elementary reactions). Then from \eqref{dyn} it can be shown that the time derivative of some statistical moments of order $l$ would now depend on moments of order higher than $l$. For example the time derivative of moments of order 1 (the means) depend on the moments of order $2$ if the propensity functions are quadratic. Thus for nonlinear propensities, the time evolution of the vector $\boldmath{\mu}$ consisting of all the moments up to order ${{L}}$ of $\x$ is given by:
\begin{equation}\label{eq:linearsysu}
\small
\frac{d{\boldmath{\mu}}}{dt}={a}+A {{{ \boldmath{\mu}}}}+ B \tilde{ \boldmath{\mu}},
\end{equation}
for an appropriate constant vector  ${a}$, constant matrices $A$ and $B$, and a (time-varying) vector $\tilde{\boldmath{\mu}}$ containing moments of order ${{L}}+1$ and higher \cite{Singh:2011IEEETAC}. We illustrate this point using an example of negative auto-regulation,
where a protein inhibits its own transcription. Such systems are common motifs in gene networks and play critical roles in controlling protein noise levels  \cite{Alon:2007NatureReviewsGenetics,Kepler:2001,Singh:2009B,Becskei:2000,Simpson:2003PNAS,Nevozhay:2009,Singh:2011IEEETN,Dublanche:2006}. Consider the reaction set:
\begin{equation}\label{re}
\begin{aligned}
&Gene_{OFF} \xrightarrow{~k_{on}~} Gene_{ON}, \  Gene_{ON} \xrightarrow{~k_{off}x^2_2~} Gene_{OFF}, \\ & Gene_{ON} \xrightarrow{~k_p
~} Gene_{ON} + U \times  Protein, \ Protein \xrightarrow{~\gamma_p~} \emptyset,
\end{aligned}
\end{equation}
where the gene switches between an active (ON) and inactive (OFF) state. Let $x_1(t)$ be a Bernoulli random variable with $x_1(t)=1$ ($0$) denoting that the gene is active (inactive) at time $t$.  Assuming mRNA half-life is considerably shorter than the protein half-life,
we ignore mRNA dynamics and model protein synthesis directly from the ON state in geometric bursts of size $U$ \cite{Shahrezaei:2008PNAS,Golding:2005Cell,Cai:2006Nature}. 
A negative feedback is implemented by making the gene inactivation rate $k_{off} x_2^2$ increase quadratically with the protein level $x_2(t)$; this effectively models the scenario where two proteins cooperatively bind to the promoter to turn it off. This implies a nonlinear propensity function $k_{off}x_2^2 x_1$ for the ON to OFF reaction in the stochastic model \cite{Soltani:2015IEEETBSC}. Note that typically when the gene activates back, two protein molecules should be released but for simplicity here we assume that the protein numbers are so high that we can ignore the latter, i.e. in our model the only processes which affect protein numbers are translation and degradation.
It can then be deduced using \eqref{dyn} that the time evolution of a vector $\boldmath{\mu}$ consisting of the first, second and third moments of $x_1(t)$ and $x_2(t)$ is given by: 
\begin{equation}
\label{ho}
\small
\frac{d{\boldmath{\mu}}}{dt}={a}+A {{{ \boldmath{\mu}}}}+ B \tilde{ \boldmath{\mu}}, \ \ \tilde{\boldmath{\mu}}=[\langle x_1x_2^3 \rangle, \langle x_1x_2^4 \rangle ]^T
\end{equation}
and depends on fourth and fifth-order moments \cite{Soltani:2015IEEETBSC}. {\em This example illustrates the general principle that nonlinear propensities result in non-closed moment dynamics, where time evolution of lower-order moments depends on higher-order moments}. The lack of closure for these systems presents a significant challenge in computing statistical moments for nonlinear biochemical processes. Next we discuss two different approximate methods, namely moment-closure and the Linear-Noise approximation, for approximately solving the moment equations.

\section{Moment closure methods}
 One way to solve systems of the form \eqref{eq:linearsysu} is to close moment dynamics by approximating higher-order moments $\tilde{\boldmath{\mu}}$ as nonlinear functions of lower-order moments in $\boldmath{\mu}$, as in $\tilde{\boldmath{\mu}}\approx \varphi(\boldmath{\mu})$. This procedure is called \emph{moment closure} \cite{Whittle:1957JRSSB,Krishnarajah:2005BuB,Konkoli:2012JTB,Smadbeck:2013PNAS,Grima:2012JCHP} and results in nonlinear approximated moment dynamics of the form:  
 \begin{equation}\label{eq:truncated}
 \small
\frac{d{\boldmath{\nu}}}{dt} = {a}+A {{{\boldmath{\nu}}}} + B { \varphi}({{{\boldmath{\nu}}}}),
 \end{equation} 
where the state of this closed system $\boldmath{\nu}(t)$ can be  viewed as an approximation for $\boldmath{\mu}(t)$. 

If we want an approximate closed set of moment equations for all the moments up to order $L$, then the standard approach for obtaining functions $\varphi$ is to assume that the $L+1$ and higher-order moments come from a certain probability distribution. This assumption is of course ad-hoc but practical and simple to implement. The most commonly used closure method obtains approximate equations for the first two moments by assuming that the third cumulant is zero (often referred to as the two-moment approximation). This is consistent with the assumption of a multivariate Gaussian distribution since the skewness is zero \cite{GomezUribe:2007,Ullah:2009TB,Matis:2002EES,Gillespie:2009IET,Grima:2012JCHP}. The closed system of equations for the two-moment approximation for a system with nonlinear propensities may become unstable at low population sizes resulting in moment estimates growing unboundedly or becoming negative or even leading to incorrect oscillatory dynamics \cite{Azunre:2011IET,Nasell:2003,Singh:2011IEEETAC,Schnoer:2014JCHP}.

Some of the deficiencies of this method can be corrected by deriving equations for all the moments up to order $L$ (where $L >= 3$) and then assume the $L+1$ cumulant to be zero. This $L$-moment approximation does not assume zero skewness as the two-moment approximation and hence it can account for a wide variety of non-Gaussian distributions. Indeed it has been shown to accurately describe a number of chemical systems with moderately small copy numbers \cite{Grima:2012JCHP}. Generally in the limit of large system sizes (or equivalently of molecule numbers), it can be shown that the accuracy of the moment approximation for deterministically monostable systems increases with $L$; to be more precise, the absolute error between the mean concentration and the variance of fluctuations about them, calculated using the $L$-moment approximation and the exact one, is proportional to $\Omega^{-L}$ \cite{Grima:2012JCHP}, where $\Omega$ is the system size, i.e., the volure of the system. It has also been shown that these approximations generally give physically meaningful results (such as positive mean and variance, a single-solution for the steady-state moments and no unphysical oscillations) provided the system-size is larger than a critical threshold and for deterministically bistable or oscillatory systems only if the system-size is bounded from above and below \cite{Schnoer:2014JCHP}. Other types of closures are based on an assumption of the Gamma, log-normal or Poisson distributions or else by neglecting the central moments (also called low dispersion closures). These often also lead to improvements over the two-moment Gaussian approximation for certain parameter values; for a detailed comparison between different types of moment closure methods see for example \cite{Keeling:2000JTB,Schnoer:2015JCHP,Lakatos:2015JCHP,Singh:2007ACC}.

A variation on the above methods is the recently introduced method of conditional moments \cite{Hasenauer:2014JMB}. The key idea is to use a CME description for low-copy number species and a moment-based description for medium or high-copy number species. This is done by conditioning the moments of the abundant species on the state of the low abundance species. This technique is particularly useful for the modelling of stochastic gene expression where a natural abundance separation exists between the number of promoters, mRNA (low abundant species) and the number of proteins (highly abundant species). For example, many mRNA species in \emph{E. coli} and yeast are present at an average of 1 molecule or less per cell \cite{Taniguchi:2010Science,Bar-Even:2006NatureGenetics,Newman:2006NatureGenetics}. Moreover, genes can randomly transition between active and inactive states, and the number of active copies of a gene can be zero with high probability \cite{Suter:2011Science,Brown:2013PLOSBiology,Raj:2006PLOSBiology,Hornung:2012GenomeResearch,Corrigan:2014CurrentBiology,Bothma:2014PNAS,Chubb:2006Currentbiology,Daigle:2015Bioinformatics}. For these applications the method of conditional moments leads to considerably more accurate estimates of the moments compared to the conventional closure schemes described above. Recently, it has been shown that accurate approximations for the pdf of the CME can be calculated by using the moments obtained from this closure scheme together with the maximum entropy principle \cite{Andre:2015arxiv}. 

A different set of closure techniques are based on \emph{derivative-matching}, where the moment closure is done by matching time derivatives of the exact (not closed) moment equations \eqref{eq:linearsysu} with that of the approximate (closed) moment equations \eqref{eq:truncated} at some initial time $t_0$ \cite{Singh:2007BuB,Singh:2011IEEETAC,Singh:2010PTRSA}. These are different than the closure methods discussed earlier since they do not specifically assume a distribution to close the moment equations.  
More specifically, this derivative-matching approach attempts to determine nonlinear functions $\varphi$ in \eqref{eq:truncated} for which 
\begin{equation}\label{match}
\small
 \frac{d^i{\boldmath{\mu}}(t)}{dt^i}\big|_{t=t_0}=\frac{d^i {{{\boldmath{\nu}}}}(t)}{dt^i}\big|_{t=t_0}
 \end{equation}
holds for deterministic initial conditions (where the mean is equal to that of the deterministic chemical rate equations and the second and higher-order central moments are zero). The main rationale for doing so is that, if a sufficiently large number of derivatives of $\boldmath{\mu}(t )$ and $\boldmath{\nu}(t )$ match point-wise at an initial time $t_0$, then from a Taylor series argument the trajectories of $\boldmath{\mu}(t )$ and $\boldmath{\nu}(t )$ will remain close at least locally in time.  Interestingly, analysis reveals that a class of functions $\varphi$ for which equation \eqref{match} holds approximately for all $i\geq 1$ (i.e., all time derivatives of $\boldmath{\nu}(t)$ and $\boldmath{\mu}(t)$ match at  $t=t_0$ with small errors) indeed exists \cite{Singh:2011IEEETAC}. Theorem 1 in \cite{Singh:2011IEEETAC} provides explicit formulas to construct moment closure functions $\varphi$ expressing any arbitrary higher-order moment in terms of lower-order moments. For example, based on this derivative-matching technique, fourth and fifth-order moments in \eqref{ho} can be approximated in terms of moments up to order three as follows
\begin{equation}\label{dm_3}
\small
 \langle x_1 x_2^3 \rangle \approx\left(\frac{\langle x_1 x_2^2 \rangle}{\langle x_1 x_2 \rangle}\right)^3 \left(\frac{\langle x_2 \rangle}{\langle x_2^2 \rangle}\right)^3 \langle x_2^3 \rangle  \langle x_1 \rangle,
 \quad
\langle x_1 x_2 ^4\rangle \approx\frac{\langle x_1 \rangle^3  \langle x_2 \rangle^{12}   \langle x_2^3 \rangle^{4}\langle x_1 x_2^2 \rangle^6 }{\langle x_2^2 \rangle^{12} \langle x_1 x_2 \rangle^8}
\end{equation}
\cite{Singh:2011IEEETAC,Soltani:2015IEEETBSC}. 
A feature of the derivative-matching closure technique is that if we close the dynamics of all the moments up to order ${{L}}$, then the error in matching derivatives in \eqref{match} decreases as $\Omega^{-L}$, where $\Omega$ is a measure of system size \cite{Singh:2011IEEETAC}. Thus by increasing ${{{L}}}$, which corresponds to including higher-order moments in the  vector $\boldmath{\mu}$,  the approximated moment dynamics \eqref{eq:truncated} provides more accurate approximations to the exact moment dynamics \eqref{eq:linearsysu}. 

\begin{figure}[!t]
	\centering
	\includegraphics[width=3in, angle =90]{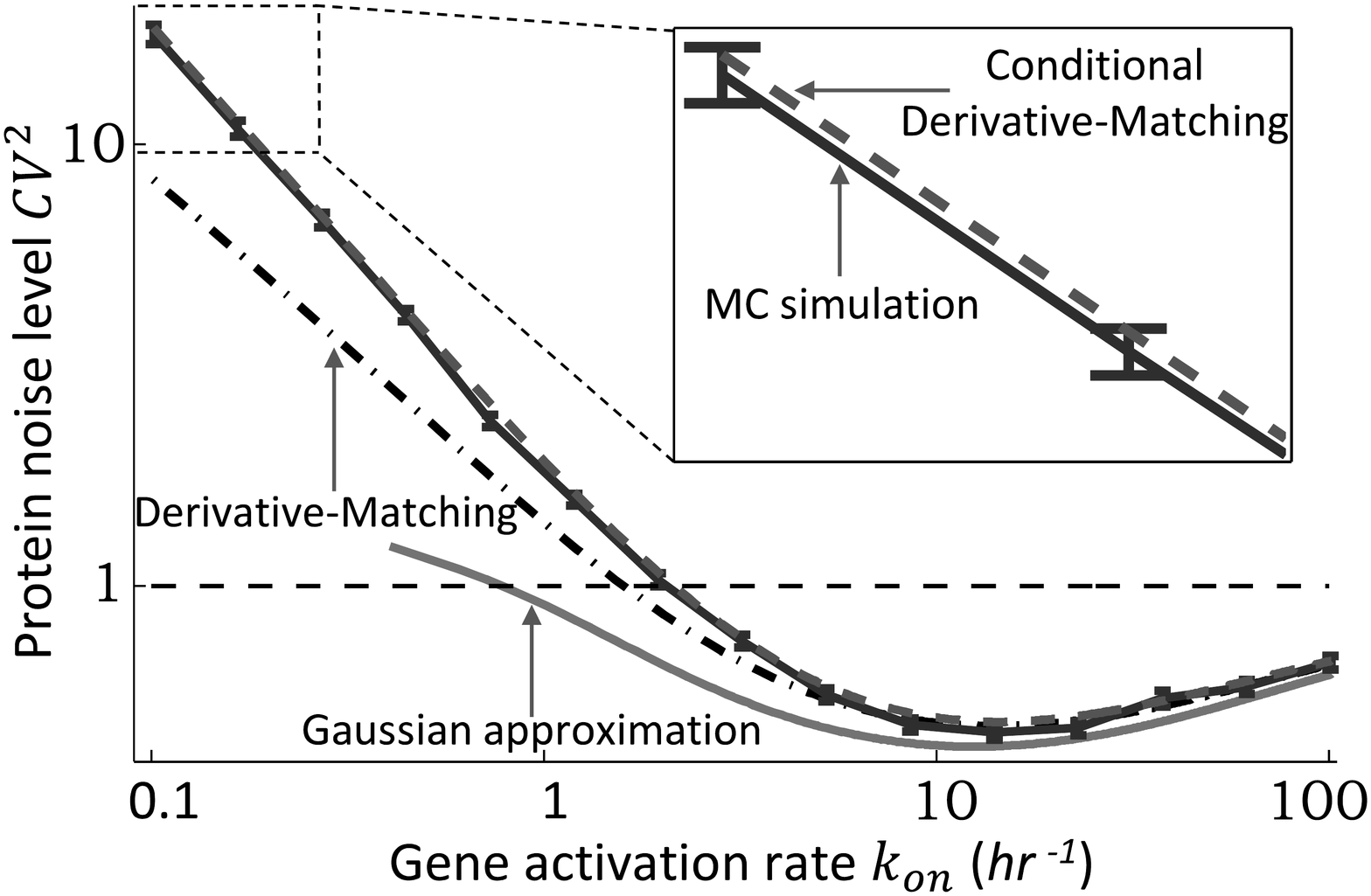}
	\caption{\small  {\bf A comparison of Conditional Derivative-Matching (CDM), Derivative-Matching (DM) and the two-moment (Gaussian) approximation}. Steady-state protein noise levels as a function of $k_{on}$ in \eqref{re} for different closure techniques. Mean protein level is fixed at $200$ molecules by simultaneously changing $k_p$. 
		Exact noise levels obtained by running  $10,000$ Monte Carlo (MC) simulations using the SSA. Inset shows the match between estimates from CDM and MC simulations. Error bars represent 95\%
		confidence intervals computed using bootstrapping. Noise levels are normalized by their corresponding value when there is no feedback. See main text for the details of CDM, and DM methods used. Parameters taken as $\langle U \rangle=70$, $\gamma_p=1 \ hr^{-1}$ and  $k_{off}=10^{-3} \ hr^{-1}$.  }
	\label{SinghFig2}
\end{figure}

As for traditional moment closure methods, the derivative-matching closure technique \cite{Singh:2011IEEETAC} does not do well in cases where some species have very low copy numbers, as often the case in cellular processes.  To deal with these conditions one can employ the Conditional Derivative-Matching (CDM) closure technique \cite{Soltani:2015IEEETBSC}. This works in much the same way as the method of conditional moments described earlier except that here the higher-order conditional moments are expressed in terms of lower-order conditional moments as per the derivative-matching closure technique. We illustrate this technique using the auto-regulation example \eqref{re}, where the goal is to approximate  higher-order moments $\langle x_1 x_2^3 \rangle, \langle x_1x_2^4 \rangle$ as functions of moments up to order three. Since the gene's transcriptional status $x_1(t) \in \{0,1\}$ is binary, moments are first conditioned on $x_1(t)=1$ (gene being active) 
\begin{equation}\label{cond}
\small
\langle x_1x^l_2 \rangle =\langle x^l_2 | x_1=1 \rangle{\rm Probability}(x_1=1)=\langle x^l_2 | x_1=1 \rangle\langle x_1 \rangle, \ \ l\in\{1,2,\ldots\}.
\end{equation}
Next, higher-order conditional moment $\langle x^3_2\vert x_1=1 \rangle$ is expressed in terms of lower-order conditional moments as per the derivative-matching closure technique. 
Based on derivative-matching, the third-order moment of any random variable $y$ can be approximated as {${\langle y^3 \rangle\approx {\langle y^2 \rangle^3}/{\langle y \rangle^3}}$} \cite{Singh:2011IEEETAC}. Using this result and \eqref{cond} yields the following approximation for $\langle x_1x_2^3 \rangle$ in terms of moments up to order three
\begin{align}\label{cdm1}
\small
\langle x^3_2\vert x_1=1 \rangle \approx \frac{\langle x^2_2 |  x_1=1  \rangle^3}{ \langle x_2 |  x_1=1  \rangle^3} \implies  \langle x^3_2\vert x_1=1 \rangle\langle x_1 \rangle &\approx \frac{\left(\langle x^2_2 |  x_1=1  \rangle\langle x_1 \rangle\right)^3}{\left( \langle x_2 |  x_1=1  \rangle\langle x_1 \rangle\right)^3}\langle x_1 \rangle \notag \\ &\implies  \langle x_1 x_2^3 \rangle \approx \left(  \frac{\langle  x_1 x_2^2 \rangle}{\langle x_1 x_2 \rangle} \right)^3 \langle x_1 \rangle.
\end{align}
A similar procedure results in 
$\langle x_1x_2^4 \rangle \approx  {\langle x_1  \rangle^3  \langle x_1 x_2^2  \rangle^6}/ {\langle x_1 x_2  \rangle^8}$ \cite{Soltani:2015IEEETBSC}.
Fig. \ref{SinghFig2} reveals that protein noise levels, as measured by the coefficient of variation squared  ($CV^2 = {\text{variance}}/{\text{mean}^2}$), 
obtained from CDM are remarkably close to their exact values, even in regimes of slow gene switching ($k_{on} \ll \gamma_p$ in \eqref{re}), where the protein population counts have a bimodal distribution. Conditional moments can in principle be used to reconstruct the bimodal distribution based on a mixture of conditional distributions. Interestingly, noise is minimized at an intermediate value of $k_{on}$, suggesting an optimal protein binding affinity provides the best control of random fluctuations.

{In summary, a wide variety of moment closure approximations exist. These provide exact moments when the propensities are linear (since in this case no closure is necessary) but typically provide approximate moments when the propensities are nonlinear. The accuracy of most approximations decreases with the average molecule numbers, i.e. they are often useful in cases where the fluctuations are not large compared to the mean molecule numbers. While closure techniques can provide good approximations of the moment dynamics, there are no mathematical guarantees on the accuracy of the approximation. This motivates development of techniques that can bound the actual moment dynamics, even though the stochastic system might be analytically intractable, and we refer interested readers to \cite{Ghusinga:2016arXiv,Lamperski:2016CDC} for progress in this direction.}

\section{The Linear-Noise Approximation}

Another way to approximate the solution to the moment dynamics equation \eqref{eq:linearsysu} is to use the \index{linear! noise approximation}Linear-Noise Approximation (LNA) \cite{vanKampen,Elf:2003}. This method constitutes a small noise approximation of the probability distribution solution of the CME; this leads to a Gaussian distribution \index{Gaussian! distribution}centered about the mean molecule number predicted by the chemical rate equations. It agrees with the two-moment approximation in the macroscopic limit of large volumes \cite{Grima:2012JCHP}, namely in the limit of small noise. The advantages of this method are numerous: (i) the mean and variance are always guaranteed to be positive and hence physically meaningful (unlike moment-closure approximations); (ii) the first two moments are exact for systems composed of at most first-order reactions and also for a small class of systems with bimolecular reactions \cite{Grima:2015PRE}; (iii) the method provides an approximation for most systems with bimolecular reactions, but the error (relative to the exact CME solution) decreases with increasing molecule numbers. In particular the error in the mean concentrations and the variance about them is roughly proportional to the inverse mean number of molecules and the inverse mean number of molecules squared, respectively \cite{Grima:2010JCHP,Grima:2011JCHP} (or equivalently the inverse volume and the inverse volume squared respectively); (iv) dimensional reduction due to timescale separation is much easier to perform on the LNA compared to the CME and this allows one to describe complex systems with just few variables \cite{Thomas:2012BMC}; (v) the LNA is obtained by solving a system of linear equations and hence avoids the numerical problems encountered when solving the nonlinear equations obtained by moment-closure methods. A major limitation is that it can describe only systems characterized by a unimodal distribution; however a recent extension (called the conditional LNA) enables its use to approximate systems which possess a multimodal distribution due to slow promoter switching \cite{Thomas2014PNAS}. 

In what follows we present a sketch of van Kampen's derivation \cite{vanKampen} of the LNA of the \index{chemical master equation}chemical master equation  (\ref{CMEgen}) (see also the one in \cite{Elf:2003}). Let us first consider the macroscopic limit of the CME. Specifically, this is defined as the limit of large volume at constant concentration or in other words the limit of large molecule numbers. Now by invoking the Central Limit Theorem\index{Central Limit Theorem}  and the Law of Large Numbers\index{Law of Large Numbers}, we can say that the standard deviation of noise roughly scales as the square root of the mean number of molecules. Thus in the macroscopic limit of large molecule numbers, the size of the fluctuations is small compared to the mean number of molecules, i.e., the macroscopic limit is also the deterministic limit. Now we know from \emph{in vitro} experiments that in this limit, the mean molecule numbers are accurately described by the standard ordinary differential equations (i.e., the chemical rate equations discussed in Chapter 3). Specifically, the chemical rate equations are equations for the concentration vector ${\bf{c}}=(c_1,..,c_N)^T$ where $c_i$ is the concentration of species $S_i$. By multiplying this concentration by the volume $\Omega$ of the system, one obtains the vector of the mean number of molecules $\Omega \bf{c}$. Thus, we come to the conclusion that in the macroscopic limit,  we expect the marginal distribution solution of the CME of species $S_i$ to be sharply peaked around the rate equation solution $\Omega c_i$ with a width of {\emph{order}} of the square root of the molecule numbers, i.e. of order $\Omega^{1/2}$. Mathematically this assumption can be written as: 
\begin{equation}
\label{VKansatz}
x_i = \Omega c_i + \Omega^{1/2} e_i, 
\end{equation}
where $e_i$ is a new fluctuating variable. The term $\Omega^{1/2} e_i$ is the noise about the deterministic mean number of molecules. The idea then is to find what is the distribution $\Pi({\bf{e}},t)$ of the vector of fluctuations ${\bf{e}}=(e_1,...,e_N)^T$ at time $t$. First we write the chemical rate equations as:
\begin{equation}
\label{RE}
\frac{d}{dt} c_i = \sum_{j=1}^M \psi_{ij} (\bf{c}),
\end{equation}
where $\psi_{ir} (\bf{c})$ is the term in the rate equation which describes how reaction $R_j$ affects the concentration of species $S_i$.  Then substituting \eqref{VKansatz} in the CME \eqref{CMEgen}, one can show that the leading-order term of a Taylor series expansion of the CME in powers of $\Omega^{-1/2}$ \cite{vanKampen,Elf:2003,Grima:2010JCHP}  is given by a Fokker-Planck equation\index{Fokker-Planck equation}:
\begin{equation}
\label{FPE}
\frac{\partial \Pi({{\bf{e}},t)}}{\partial t} = -\sum_{i,k=1}^N J_{ik} \frac{\partial}{\partial \epsilon_i} (\epsilon_k \Pi({\bf{e}},t)) + \frac{1}{2} \sum_{i,k=1}^N D_{ik} \frac{\partial^2}{\partial \epsilon_i \epsilon_k} \Pi({\bf{e}},t),
\end{equation}
where
\begin{align}
J_{ik}= \frac{\partial}{\partial c_k} \sum_{j=1}^M \psi_{ij} ({\bf{c}}), \ \ \
D_{ik} = \sum_{j=1}^M \psi_{ij} ({\bf{c}}) S_{kj}.
\label{matrixdefs}
\end{align}
We remind the reader that the stoichiometry\index{stoichiometry} $S_{kj}$ is the net change in the number of molecules of species $S_k$ when reaction $R_j$ occurs. The chemical rate equations \eqref{RE} together with the Fokker-Planck equation \eqref{FPE}\index{Fokker-Planck equation} constitute the LNA; the former approximates the means of the molecule numbers and the latter gives a description of the noise about these means. 

From the Fokker-Planck equation\index{Fokker-Planck equation}, it is possible to calculate the moments of the fluctuation vector $\bf{e}$ at all times. If one starts with deterministic initial conditions, i.e., second and all the higher order centred moments are zero initially, then it can be shown that $\langle e_i \rangle = 0$ at all times. The second-moments $C_{ij} = \langle e_i e_j \rangle$ are zero initially but become non-zero as time progresses. The time-evolution equation for these moments is given by the Lyapunov equation\index{Lyapunov! equation}:
\begin{equation}
\frac{d}{dt} C_{ik} =  \sum_{j=1}^M (J_{ij} C_{jk}+ C_{ij}  J_{kj})+D_{ik}.
\label{Leq}
\end{equation}
By the assumption \eqref{VKansatz}, it follows that the covariance of fluctuations in the number of molecules of species $S_i$ and $S_k$ is given by $\Omega C_{ik}$. Hence, in summary, the LNA involves solving the chemical rate equations \eqref{RE} and the Lyapunov equation \eqref{Leq}.  

We now demonstrate the use of the LNA by means of an example. We consider the Michaelis-Menten reaction with substrate input \cite{Grima:2009BMC}:
\begin{align}
\null \xrightarrow{k_{in}} S, \ \ \ \
S + E  \xrightarrow{k_0} C, \ \ \ \
C  \xrightarrow{k_1} S + E, \ \ \ \
C  \xrightarrow{k_2} E + P,
\label{MM}
\end{align} 
where $S$, $E$, $C$ and $P$ denote substrate, free enzyme, complex and product species and the $k$'s denote the associated rate constants in the chemical rate equations. The first reaction models the translation of substrate molecules, while the subsequent reactions involving enzyme or complex, model the catalysis of substrate into product molecules. We shall apply the LNA to obtain approximate expressions for the mean number of molecules and the variance of fluctuations about these means for the substrate and enzyme species. Note that we shall not be concerned with the product; the analysis is independent of the statistics of product numbers since the product is produced irreversibly and does not participate in any reaction.

To compute the LNA we need two pieces of information. The individual contributions of each reaction to the dynamics of species, i.e., the elements $\psi_{ij}$, and the stoichiometric matrix elements $S_{ij}$. From the chemical rate equations, by inspection, one can easily deduce that the individual contributions of each reaction to the dynamics of species $S_1$ are: $\psi_{11} = k_{in}$, $\psi_{12} = - k_0 c_1 c_2$, $\psi_{13} = k_1 (E_T - c_2)$ and $\psi_{14} = 0$. Similarly for species $S_2$ we have: $\psi_{21} = 0$, $\psi_{22} = - k_0 c_1 c_2$, $\psi_{23} = k_1 (E_T - c_2)$ and $\psi_{24} =  k_2 (E_T - c_2)$, where $E_T$ is the total enzyme concentration (a constant at all times). By inspection of the mechanism \eqref{MM}, and recalling that $S_{ij}$ is the net change in the number of molecules of species $i$ when reaction $R_j$ occurs, one finds: $S_{11} = S_{13} = S_{23} = S_{24} = 1$, $S_{12} = S_{22} = -1$, and $S_{14} = S_{21} = 0$. 

Let the reactions in the reaction scheme \eqref{MM} be numbered $1$ to $4$ (left to right). Let the concentration of substrate and free enzyme be denoted as $c_1$ and $c_2$. Then from mass-action kinetics, the chemical rate equations are given by:
\begin{align}
\label{MMRe}
\frac{d}{dt} c_1 &= \sum_{j=1}^4 \psi_{1j} = k_{in} - k_0 c_1 c_2 + k_1 (E_T - c_2) + 0, \nonumber \\
\frac{d}{dt} c_2 &= \sum_{j=1}^4 \psi_{2j} = 0 - k_0 c_1 c_2 + k_1 (E_T - c_2) + k_2 (E_T - c_2).
\end{align}
Note that each term corresponds to a reaction and a zero implies no contribution of a reaction to the dynamics of a particular species. Since enzyme molecules are either in the free state $E$ or in the complex state $C$, it follows that the number of $C$ molecules is given by $E_T - c_2$ (this conservation law has been used in the chemical rate equations above). 

Now we have all the information to derive the LNA. We shall for simplicity specify steady-state conditions, as this simplifies the algebra. Solving the chemical rate equations \eqref{MMRe} in steady-state conditions (setting the time derivative to zero), we obtain:
\begin{align}
c_1 &= \frac{K_M \Lambda}{1 - \Lambda}, \label{smean} \\
c_2 &= E_T (1 - \Lambda),
\label{emean}
\end{align}
where $K_M=(k_1 + k_2)/k_0$ (this is the Michaelis-Menten constant) and $\Lambda = k_{in} / k_2 E_T$. Note that a steady-state only exists if $\Lambda < 1$; this is since there is a finite maximum reaction rate at which the enzyme can catalyse substrate into product and if the rate of substrate input exceeds this maximum, i.e. if $\Lambda > 1$, then the substrate will increase indefinitely with time (the enzyme number is fixed by the conservation law and so the non-equilibrium condition is only for substrate and product). 

\begin{figure}[!t]
\centering
\includegraphics[width=4.5in]{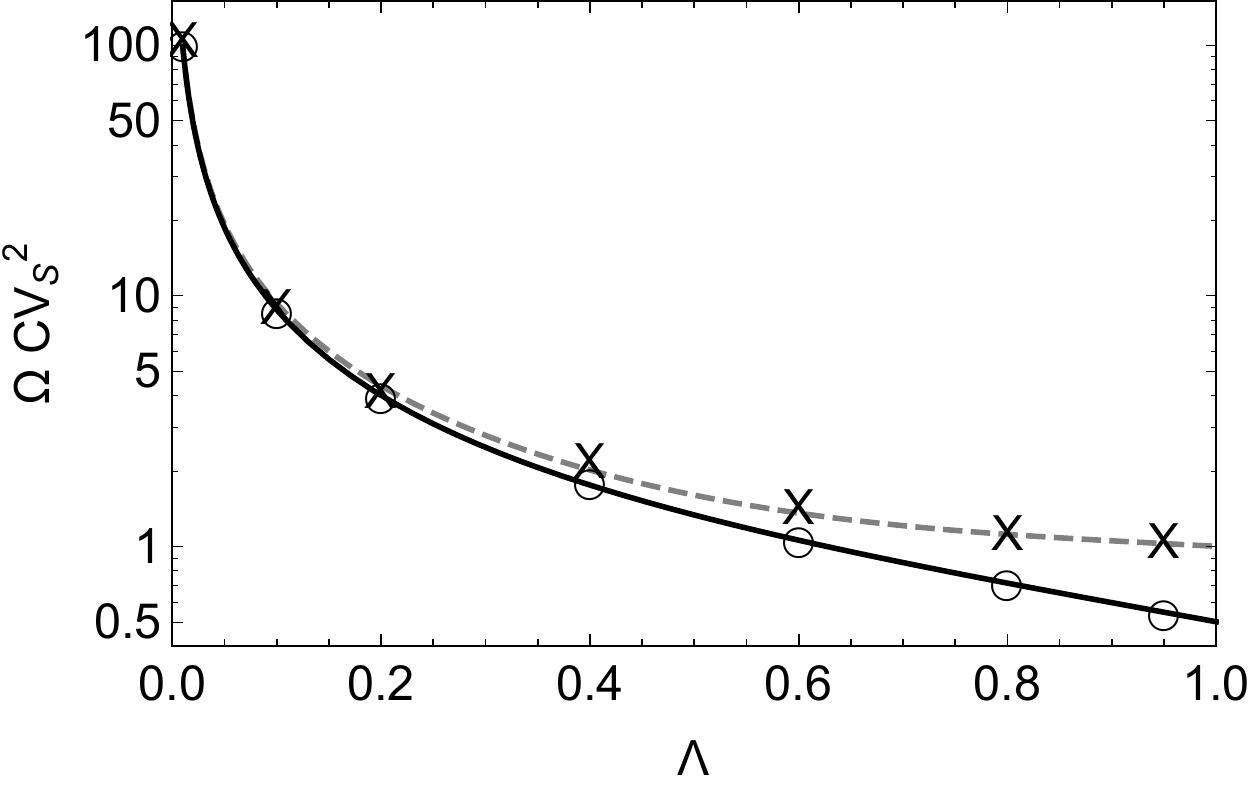}
\caption{\small  {\bf  Testing the accuracy of the LNA's prediction for the size of substrate fluctuations versus stochastic simulations in steady-state conditions}. The LNA prediction (for all volumes, dashed gray line) is obtained by using \eqref{smean} and \eqref{sflucs} to calculate the coefficient of variation squared of substrate number fluctuations $CV_S^2$ as a function of $\Lambda$ while keeping $R = K_M = E_T$ fixed to one. The stochastic simulations are performed at three different volumes $\Omega = 10, 100$ (crosses) and $\Omega = 1$ (open circles). Note that the data for $\Omega = 10$ and $\Omega= 100$ are not distinguishable from each other. The solid black line shows the exact solution of the CME for the case of one enzyme molecule, i.e., for $\Omega = 1$ (this is found in Appendix G of \cite{Schnoer:2014JCHPa}). The LNA predictions agree well with stochastic simulations for all volumes except $\Omega = 1$; the discrepancy becomes worse as the point $\Lambda = 1$, at which the system switches to non-equilibrium conditions for substrate numbers, is approached.}\label{SinghFig3}
\end{figure}

\begin{figure}[!t]
\centering
\includegraphics[width=4.5in]{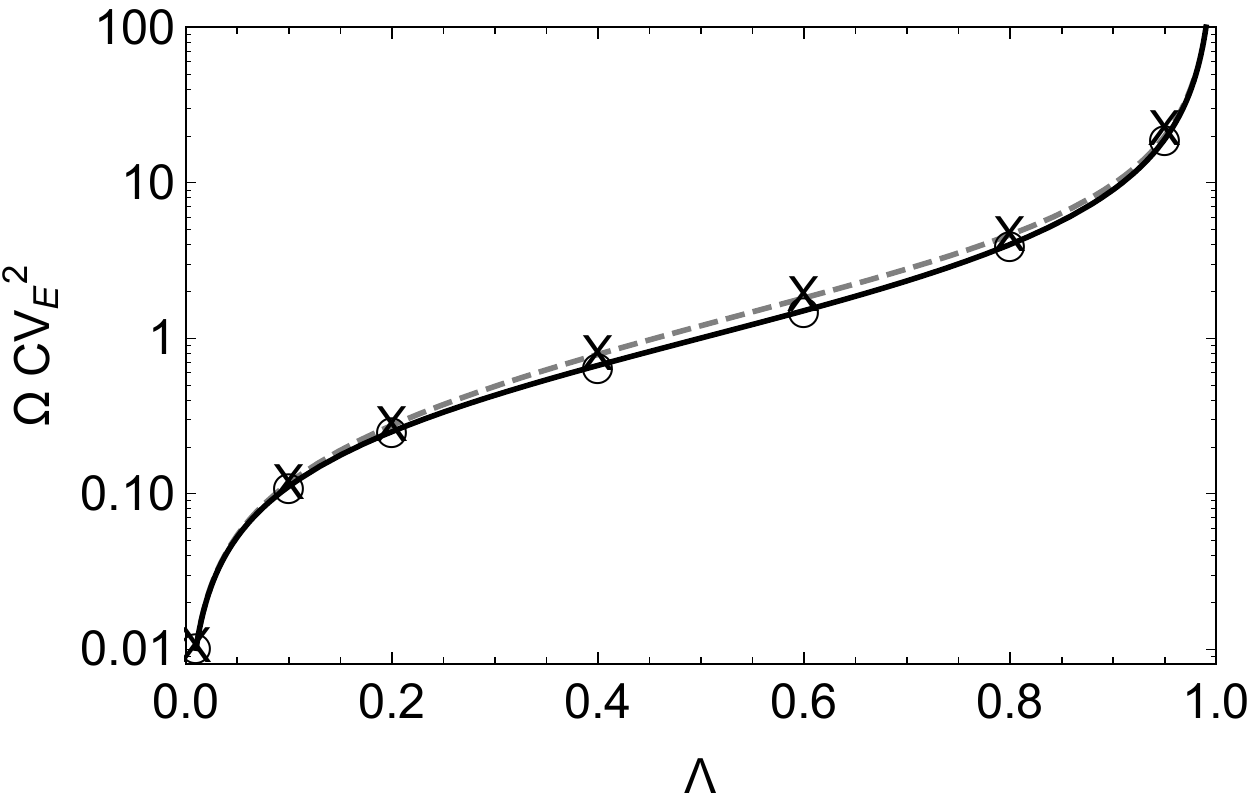}
\caption{\small  {\bf  Testing the accuracy of the LNA's prediction for the size of enzyme fluctuations versus stochastic simulations in steady-state conditions}. The LNA prediction (for all volumes, dashed gray line) is obtained by using \eqref{emean} and  \eqref{eflucs} to calculate the coefficient of variation squared of free enzyme number fluctuations  $CV_E^2$ as a function of $\Lambda$ while keeping $R = K_M = E_T$ fixed to one. The stochastic simulations are performed at three different volumes $\Omega = 10, 100$ (crosses) and $\Omega = 1$ (open circles). Note that the data for $\Omega = 10$ and $\Omega= 100$ are not distinguishable from each other. The solid black line shows the exact solution of the CME for the case of one enzyme molecule, i.e., for $\Omega = 1$ (this is found in Appendix G of \cite{Schnoer:2014JCHPa}). The LNA predictions agree well with stochastic simulations for all volumes.}\label{SinghFig4}
\end{figure}

The covariance of the fluctuations about these mean concentrations are obtained by: (i) substituting the elements $\psi_{ij}$, and the stoichiometric matrix elements $S_{ij}$ in \eqref{matrixdefs} to calculate the elements $J_{ik}$ and $D_{ik}$; (ii) substituting the latter into \eqref{Leq} with the time-derivative set to zero. This equation reduces to solving a set of linear algebraic equations for the quantities $C_{11},C_{12},C_{21},C_{22}$:
\begin{align}
C_{11} &= \frac{K_M \Lambda (-E_T (\Lambda - 1)^3 (1 + R) + K_M (1 + (\Lambda - 1) \Lambda + R))}{(K_M + E_T (\Lambda - 1)^2) (\Lambda - 1)^2 (1 + R)}, \label{sflucs} \\
C_{12} &= C_{21} = -\frac{ E_T K_M \Lambda^2}{K_M + E_T (\Lambda - 1)^2}, \\
C_{22} & = \frac{E_T (-K_M + E_T (\Lambda - 1)) (\Lambda - 1) \Lambda}{K_M + E_T (\Lambda - 1)^2},
\label{eflucs}
\end{align}
where $R = k_1 / k_2$. For the substrate fluctuations, the coefficient of variation squared (variance of molecule numbers divided by the square of the mean molecule numbers) is given by $CV^2_S = \Omega C_{11} / (\Omega c_1)^2 = C_{11} / \Omega c_1^2$ while for the enzyme we correspondingly have $CV_E^2 = C_{22} / \Omega c_2^2$. Note that the LNA predicts that the quantities $\Omega CV_S^2$ and $\Omega CV_E^2$ are volume independent. Notably this implies that provided the total enzyme concentration $E_T$ is kept constant, according to the LNA, the quantities $\Omega CV_S^2$ and $\Omega CV_E^2$ are independent of the number of enzyme molecules. To test the accuracy of the LNA in Figures \ref{SinghFig3} and \ref{SinghFig4} we plot the LNA predictions for these two quantities versus the non-dimensional parameter $\Lambda$ and compare with the same obtained from Monte Carlo simulations using the Gillespie algorithm. Note that the propensities used in this algorithm for the Michaelis-Menten reaction \eqref{MM} are: $w_1=k_{in} \Omega$ , $w_2 = (k_0 / \Omega) x_1 x_2$, $w_3 = k_1 (E_T \Omega - x_2)$ and $w_4 = k_2 (E_T \Omega - x_2)$, where $x_1$ and $x_2$ are the number of substrate and of free enzyme molecules respectively. Note the propensities are written in terms of the rate constants appearing in the chemical rate equations and hence the volume $\Omega$ is necessary in the propensities such that these have units of inverse time. 

The LNA prediction is shown as a dashed gray line. We fix $R = 1$, $K_M = 1$ and $E_T = 1$ and perform simulations at three different volumes $\Omega = 1, 10, 100$. These volumes correspond to 1, 10 and 100 total number of enzyme molecules, respectively. The crosses show the data for $\Omega = 10, 100$ and the open circles for $\Omega = 1$. Note that the simulations and the LNA predictions agree remarkably well for 10 and 100 total enzyme molecules; for 1 enzyme molecule, the prediction for the substrate fluctuations is visibly less accurate, particularly as $\Lambda \rightarrow 1$, i.e. as the system moves closer to the point $\Lambda = 1$ where it switches from a stable steady-state to non-equilibrium conditions for substrate molecule numbers. In contrast, for the enzyme fluctuations the LNA does well for all volumes. These discrepancies between the LNA and stochastic simulations are typical of other systems with a chemical conservation law. However it is also to be noted that given the simplicity of the LNA approach, the agreement between theory and simulations is rather good. It is also possible to obtain an analytical estimate of the accuracy of the LNA without resorting to simulations; we shall not pursue this here, the interested reader is referred to \cite{Thomas:2013BMCGenomics}.

Although the LNA might not be as accurate as some types of moment-closure approximations, one clear advantage of the LNA over the latter is that in this case it leads to compact formulae for the approximate covariances which allow insight into how the various rate constants affect the size of fluctuations. It is always possible to obtain such elegant expressions using the LNA for any system involving one or two species; hence it is often advantageous to apply timescale separation methods to dimensionally reduce the LNA to an effective two species system \cite{Thomas:2012BMC}. When one is interested in more than two species, then a compact analytical solution of the LNA is rarely possible but the Lyapunov equation \eqref{Leq} can always be straightforwardly solved by readily available numerical solvers since it simply consists of a set of linear simultaneous equations.

Of course the LNA is an approximation, and it should be borne in mind that in some cases it can be a crude one. For such examples involving gene regulatory networks see for example \cite{Thomas:2013BMCGenomics}. The LNA is the leading-order term of the system-size expansion of the CME; by considering higher-order terms of this expansion it is possible to systematically improve on the predictions of the LNA and to also predict the error in the LNA estimates. This has been shown for a number of chemical and biochemical systems (see for example \cite{Grima:2010JCHP,Ramaswamy:2012NatureCommunications,Thomas:2013BMCGenomics}). For a comparison of these higher-order approximations with moment-closure approximations the reader is referred to \cite{Grima:2012JCHP}.

\section{Conclusion}

Moment closure\index{moment! closure} and the LNA\index{linear! noise approximation} offer an alternative approach to understanding the stochastic dynamics of bimolecular circuits. The moments to all orders computed using the moment closure approach and the moments up to second-order in the LNA are exactly the same as those of the CME, if the propensities are linear. If they are nonlinear, then the solution of both methods is typically (though there are exceptions) an approximation of those of the CME\index{chemical master equation}. In the latter case, the errors in these approximations tend to increase with decreasing molecule number, i.e. when the fluctuations become large compared to the mean molecule numbers. The methods are based on numerical solution of ordinary differential equations\index{ordinary differential equation} and hence can be straightforwardly implemented in numerical computing environments such as {\sc Matlab} and {\sc Mathematica}. However numerous software packages also exist which take as input an SBML file (or another convenient description of the bimolecular circuit) and then automatically generate and solve the ordinary differential equations for the moments. Examples of such software are: {\sc MomentClosure} \cite{Gillespie:2009IET}, {\sc StochDynTools} \cite{Hespanha2008:proc}, {\sc{Moca}} \cite{Schnoer:2015JCHP} and {\sc{Cerena}} \cite{Atefeh2016:Plos1} for moment-closure approximation and {\sc Copasi} \cite{hoops2006}, {i\sc{NA}}\cite{Thomas:2013PLOS1}, {\sc{CERENA}} \cite{Atefeh2016:Plos1} and {{\sc StochSens}} \cite{Komorowski:2012Bioinformatics} for the LNA.
 More details on the approximation methods discussed here, including a thorough comparison can be found in \cite{Schnoerr:2017JPMT}. In conclusion, these methods by providing a direct estimate of the moments bypass the need for Monte Carlo simulations, and hence lead to a computationally efficient means of exploring the dynamics, across large swaths of parameter space, of a wide variety of bimolecular circuits.	

\section{Exercises}

\begin{enumerate}
	
	\item Consider the gene expression model introduced in \eqref{re0}:
	\begin{equation}\label{ex1}
	\begin{aligned}
	& Gene \xrightarrow{~k_m~} Gene + U\times mRNA, \quad mRNA \xrightarrow{~\gamma_m~} \emptyset, \\
	& mRNA \xrightarrow{~k_p
		~} mRNA + Protein,\quad Protein \xrightarrow{~\gamma_p~} \emptyset.
	\end{aligned}
	\end{equation}
Here mRNA molecules are produced in constant bursts of size $U$. Write the moment equations for this system of reactions and subsequently solve to obtain exact expression for the steady-state mean and noise in mRNA and protein levels.

\item Find the mean number of proteins using the rate equations and the variance of protein number fluctuations using the LNA method. Set all constants (except U) to 1, use these formulae to obtain numerical values as the burst size U is varied (in steps of 5) from 1 to 30. Then compare with exact simulations using the Gillespie algorithm or direct CME integration using the Finite State Projection analysis to determine the accuracy of the rate equations and LNA as a function of the burst size.  

\item Repeat Exercise 6.2 using the two-moment approximation instead of the LNA, i.e. write the moment equations for the first and second-order moments and assume that the third cumulant is zero to close the equations. Compare your results with exact stochastic simulations or a Finite State Projection analysis to obtain the accuracy of this type of moment closure as a function of the burst size. Is the two-moment approximation method more accurate than the LNA?

\item Sometimes information can be gleaned directly from the moment equations without explicit solution; here is an example. Consider the following model of an auto-negative genetic feedback loop: G$\rightarrow$ G + P with rate $k_0$, P $\rightarrow \emptyset$ with rate $k_1$, G + P $\rightarrow$ G$^*$ with rate $k_2$ and G$^*$ $\rightarrow$ G + P with rate $k_3$. Here G and G$^*$ mean unbound and bound promoter and P is protein. Write down the moment equations for the first and second moments and without applying any moment closure, use them to show that: (i) the mean number of proteins is smaller than $k_0/k_1$; (ii) the covariance between the unbound gene and the protein is greater than $-k_0/k_1$ and less than $k_3/k_2$; and (iii) the variance in protein numbers is smaller than $(k_0 + k_1)k_3 / (k_1 k_2)$.

\end{enumerate}

\bibliography{qbioBIB.bib}

\end{document}